\documentclass[aps,prb,superscriptaddress,twocolumn,tightenlines]{revtex4-1}
\usepackage{amssymb}
\usepackage{graphicx}
\usepackage{natbib}
\usepackage{epstopdf}
\usepackage{dcolumn}
\usepackage{bm}
\usepackage{makeidx}
\usepackage{enumerate}
\usepackage {graphicx}
\usepackage {amsmath}
\usepackage {amsfonts}
\usepackage {amssymb}
\usepackage {mathrsfs}
\usepackage {bbm}
\usepackage{subfigure}
\begin{document}
\title{Nematicity driven by hybridization in iron-based superconductors}
\author{Valentin Stanev}
\affiliation{Materials Science Division, Argonne National Laboratory, Argonne, Illinois 60439, USA}

\author{Peter B. Littlewood}
\affiliation{Physical Sciences and Engineering, Argonne National Laboratory, Argonne, Illinois 60439, USA}
\date{\today }

\begin{abstract}
In this paper we study an effective model for the normal state of iron-based superconductors. It has separate, but interacting itinerant and localized degrees of freedom, originating from the $d_{xz}$ and $d_{yz}$, and from $d_{xy}$ iron orbitals respectively. At low temperatures, below a mean-field phase transition, these different states condense together in an excitonic order parameter. We show that at even lower temperature, after another phase transition, this ordered state can spontaneously break the $C_4$ lattice symmetry and become nematic. We propose this mechanism as an explanation of the tendency towards nematicity observed in several iron-based compounds.

\end{abstract}
\maketitle
{\it Introduction.} 
The discovery of high-temperature superconductivity in iron-based materials \cite{LaOFeAs, physicaC, Paglione} is one of the most exciting recent developments in physics. It is almost certain that the origin of superconductivity is unconventional (i.e. driven by electron-electron interactions), and is very likely that the order parameter is unconventional as well \cite{Hirschfeld, ChubukovR}. The normal state properties of these materials, in contrast, at first seemed rather unremarkable -- a bad metal behavior at high temperatures, which can be followed by structural transition and antiferromagnetic phase at low temperatures. This simple picture was considerably complicated by the growing evidence of nematic fluctuations \cite{JCDavis, Fisher1, ZXShen1} or even long-range order \cite{Matsuda} present in some compounds at temperatures well above the antiferromagnetic and structural transitions. Furthermore, this  nematicity seems to be of  purely electronic (rather than structural) origin \cite{Fisher2}. Understanding this state and its connection to the other phases is an important milestone in the study of these materials.

Several scenarios have been proposed to explain this tendency towards nematicity \cite{Rafael3}. Some suggest that the nematicity is generated by the spin fluctuations, and is a precursor of the incipient antiferromagnetic state \cite{Sachdev, Rafael1, Rafael2}. Another class of models emphasises the propensity towards orbital order of the iron $3d$ states in different environments \cite{Kruger, CCChen, Lv2, WCL}. There is also a model which considers the nematic state a consequence of a more complicated ``hidden" order \citep{Jian}.

In this paper we suggest an alternative scenario for the nematic instability. We introduce a simple effective model of the normal state of these materials, that assumes the simultaneous existence of localized and itinerant electronic states (originating in the iron $d_{xy}$ and $d_{xz}$/$d_{yz}$ orbital states respectively). Such models have already been considered in the context of iron-based superconductors \cite{hybrid1, hybrid2, hybrid3, hybrid4}; we concentrate, however, on the mixing of these two types of states, driven by hybridization and interactions. We show that interactions create a tendency for an excitonic instability of the localised state with more itinerant bands on the neighbors: at the mean field level this generates a hybridization gap with the symmetric $d_{xz}$/$d_{yz}$ combination which physically represents a crossover to a screened orbital at low energies with no symmetry change. At lower temperatures, we find that such order parameter can spontaneously break the $C_4$ lattice symmetry, and thus describe a nematic state. This scenario emphasizes the coupling of $d_{xy}$ with $d_{xz}$ and $d_{yz}$ states, and naturally explains the recent angle-resolved photoemission spectroscopy (ARPES) data \cite{ZXShen2, DongFeng}. In addition, a hybridization gap opens at the transition, which may explain some of the ARPES \citep{HDing}, conductivity \cite{Tanatar, Wang}, optical \cite{Nakajima} and point-contact spectroscopy \cite{LGreene} data.

{\it Effective model.} To arrive at the effective model we start from a realistic tight-binding Hamiltonian for the three $t_{2g}$ iron $d$ orbitals (the $e_g$ orbitals 
do not contribute significantly
to the density of states close to the Fermi level and we neglect them), which can be written as \citep{Daghofer}
\begin{eqnarray}
\mathcal{H}= \mathcal{H}_{0} + \mathcal{H}_{int}.
\label{H1}
\end{eqnarray}
We use the one-iron-atom unit cell axes as our coordinate system (and note that $d_{xz}$ and $d_{yz}$ orbitals are degenerate). $\mathcal{H}_0$ contains terms like $\sum_{ij, \alpha \beta \sigma } t^{i j}_{\alpha \beta} d^{\dagger}_{i,\alpha \sigma} d_{j, \beta \sigma}$, with tunnelling amplitudes $t^{i j}_{\alpha \beta}$ (where $i$, $j$ and $\alpha$, $\beta$ are site and orbital indices, respectively) which include the effects of the pnictide/chalcogenide atoms. We assume that all orbitals are close to half-filling. $\mathcal{H}_{int}$ represents the $SU(2)$ invariant on-site interactions, which includes the intra- and inter-orbital density-density interaction and the Hund's coupling \cite{Daghofer}.

One key insight from the dynamical mean-field theory (DMFT) calculations is that in some of these compounds the  $d_{xy}$ orbitals experience the so-called ``kinetic frustration" due to the destructive interference of several tunneling paths for the  $d_{xy}$ electrons \cite{Kotliar}. This suppresses the $d_{i,xy}\leftrightarrow d_{j,xy}$ hopping, and makes $t^{ij}_{xy, xy}$
significantly smaller than the other tunneling amplitudes. Thus the $d_{xy}$ electrons are much more localized then the rest, and in some cases may even undergo an orbitally-selective Mott transition \citep{Medici, Yu}.

Based on this we suggest a simplified effective Hamiltonian, which nevertheless
preserves some of the key features of the realistic model described by Eq.\eqref{H1}. It consists
of four parts:
\begin{eqnarray}
\mathcal{H}_{eff}=\mathcal{H}_{c}+\mathcal{H}_{d}+ \mathcal{H}_{mix}+\mathcal{H}_{res}.
\label{H2}
\end{eqnarray}
 The first term describes bands of itinerant electrons (denoted by $c$), which originate
from the $d_{xz}/d_{yz}$ sector of the model:
\begin{align}
\mathcal{H}_{c}= \sum_{ \bf {k},\mu \sigma } \epsilon_{ \bf {k}, \mu} c^{\dagger}_{\bf {k}, \mu \sigma} c_{ \bf {k}, \mu \sigma},
\end{align}
where $\mu $ is a band index. The $c$ states are product of the mixing of the $d_{xz}$ and $d_{yz}$ orbitals.
Even after including some interaction effects, we assume that the renormalized band structure is similar to that of Ref. \onlinecite{Raghu}. Since we are interested in the low-energy physics we treat the four surfaces crossing the Fermi level as separate bands. Thus $\mu $ runs from $1$ to $4$, with $1$ and $2$ denoting the hole pockets around $\Gamma$ and $(\pi, \pi)$ points, and $3$ and $4$ the two electron pockets around $(0, \pi)$ and $(\pi, 0)$ points (in the so-called unfolded Brillouin zone). 

The second term in $\mathcal{H}_{eff}$ contains localized states (denoted by $d$) on a periodic lattice, with large on-site repulsion:
\begin{eqnarray}
\mathcal{H}_{d}=\sum_{i \sigma} E_0 d^{\dagger}_{i \sigma} d_{i \sigma} + U \sum_{i \sigma} n_{i,d \sigma} n_{i,d \sigma'},
\end{eqnarray}
where $n_{i, \sigma}$ is the density of $d$ electrons with spin $\sigma$ on site $i$.
It describes the narrow, strongly correlated $d_{xy}$ band, present in some of the pnictides and chalcogenides. The $d_{xy}$ states are taken as completely localized, 
and we assume $E_0$ to be below the Fermi level of the itinerant bands.

 $\mathcal{H}_{mix}$ describes the mixing between the two types of fermions:
\begin{eqnarray}
\mathcal{H}_{mix}=V\sum_{i} (d^{\dagger}_{i \sigma} c_{i,\mu \sigma} + c^{\dagger}_{i,\mu \sigma} d_{i \sigma})+ W \sum_{i} n_{i, d} n_{i, c},
\end{eqnarray}
where $c^{\dagger}_{i,\mu \sigma}$ creates an electron in a Wannier state of band $\mu$. One part of it is hybridization, originating from the $xy \leftrightarrow xz$ and $xy \leftrightarrow yz$ hopping terms in $\mathcal{H}_{0}$, which are not necessarily small. Note that we are considering only real on-site $V$ (since the $c$ states are a superposition of $d_{xz}$ and $d_{yz}$ orbitals from \emph{different} sites the on-site hybridization is allowed). More realistic $V$ has to be complex and have a long-range part. The other -- four-fermionic --  terms come from the interorbital interactions, and we assumed a simple density-density form for them. All terms in $\mathcal{H}_{int}$ which cannot be cast in such form we hide in $\mathcal{H}_{res}$ (``res" for residual). We neglect these interactions for the moment, but they can play an important role.

As a result of this simplifications we arrive at a model of periodic lattice of localized states embedded in a four-band ``sea" of itinerant electrons. Similar models have already been discussed in relation with the magnetism  of the iron-based  materials \cite{hybrid1, hybrid2, hybrid3, hybrid4}, however, the term $\mathcal{H}_{mix}$ has usually been omitted.
As we will show, it has important consequences and can lead to interesting effects in the paramagnetic state of these materials.

{\it Mean-field approximation}. We use the Hamiltonian of Eq. \eqref{H2} to study the normal state of iron pnictides and chalcogenides, well above the superconducting and the magnetic critical temperatures. 
Unfortunately, even this reduced $\mathcal{H}_{eff}$ cannot be solved exactly.

We consider the limit of very large $U$, in which the doubly-occupied $d$ states are effectively forbidden. To deal with this constraint we employ the slave-bosons technique \cite{Coleman} 
by the transformation $d_{i \sigma}\rightarrow \tilde{d}_{i \sigma} b^{\dagger}_i$. There is also an on-site constraint $\sum_{\sigma}\tilde{d}^{\dagger}_{i \sigma} \tilde{d}_{i \sigma} +  b^{\dagger}_i  b_i=1$, which can be enforced by a Lagrange-multiplier term $\lambda_i (n_{\tilde{d}_i}  +  b^{\dagger}_i  b_i- 1)$ in the Hamiltonian.


Without the interaction term in $\mathcal{H}_{mix}$ the effective Hamiltonian in this limit is very similar to the Periodic Anderson Model (PAM), studied extensively in relation to heavy-fermion materials \cite{Newns}. The mean-field treatment of the problem leads to qualitatively correct result for these compounds (with some important caveats \cite{Read, Auerbach, Millis}). Because of this analogy we expect at low temperatures the $d$ electrons in our model to delocalize due to the mixing with the itinerant bands through $V$ (there are evidence of such crossover in the normal state of iron chalcogenides \cite{Tranquada, ZXShen3}). This delocalization is described on a mean-field level by a phase transition, and the appearance a non-vanishing expectation value of $\langle b\rangle \equiv b$, and can significantly change the electronic spectrum by opening a hybridization gap.

To demonstrate this we introduce two mean fields -- $b$ and $\lambda$. $b$ describes coherent tunnelling of $d$ electrons between neighbouring sites and $\lambda$ enforces the no-double occupancy constrain (but only on average). However, we still have to deal with the four-fermion terms in $\mathcal{H}_{mix}$. We can decouple them by introducing additional mean fields in the particle-hole channel $ \phi_{i \mu}\equiv \langle \tilde{d}^{\dagger}_{i \sigma} c_{i \sigma} \rangle$. Assuming spatially uniform case we use the replacement
\begin{equation}
W \sum_{i} n_{d, i} n_{c, i}\longrightarrow -W \left( \sum_{\sigma} \phi_{\mu}  c^{\dagger}_{i, \mu \sigma} \tilde{d}_{i \sigma} - \phi^2_{\mu}\right).
\end{equation}

It is important to note that $\phi_{\mu}$ should be thought of as components of a four-dimensional vector, rather than four independent order parameters. The reason is the obvious constraint that one $d$ state can only condense together with a single $c$ state or a particular linear combination of the different $c_{\mu}$ (which one is determined by the particulars of the model). However, the remaining three orthogonal linear combinations cannot condense for the lack of available $d$ electrons (similar physics is behind the so-called ``hidden metalicity" in the three-band model of Ref. \onlinecite{Eremin}).

{\it Hybridization gap.}
We first consider a fully symmetric version of our model, with two identical circular hole bands around $\Gamma$ and $(\pi, \pi)$ points, and two identical circular electron bands around $(\pi, 0)$ and $(0, \pi)$ points. For simplicity we assume that $\phi_1=\phi_2=0$, i.e. no hybridization of the $d$ electrons with the hole bands (adding these fields is straightforward). The fact that the two electron bands are  identical seems to imply that the order parameter can be freely rotated in the $(3,4)$ subspace. However, as shown in the next section, we have to take \emph{symmetric} combination of $\phi_{1}$ and $\phi_{2}$, or using vector notation $(0,0,\phi,\phi)$. Note that this order parameter preserves the underlying lattice $C_4$ symmetry (due to the $3 \leftrightarrow 4$ i.e. $x \leftrightarrow y$ symmetry of the order parameter). Also, as already explained, it leaves the other three ``bands" -- $c_1$, $c_2$ and the antisymmetric combination of $c_3$ and $c_4$ -- decoupled on mean-field level.

\begin{figure}[h]
\includegraphics[width=0.45\textwidth]{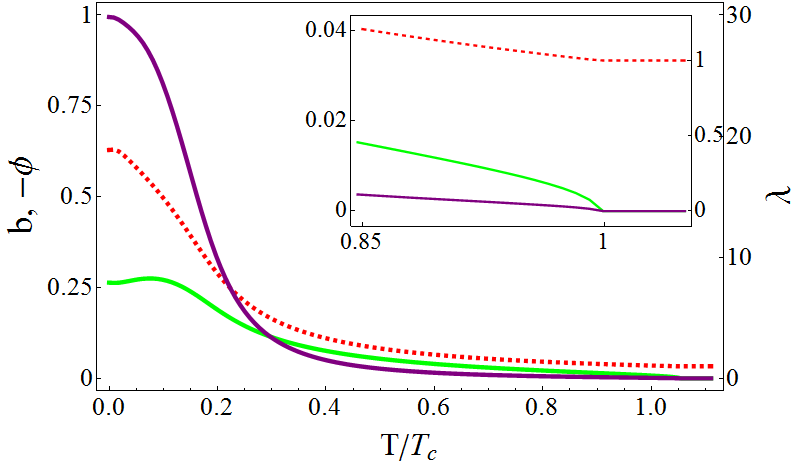}
\caption{Plot of the solutions for $\phi$ (purple), $b$ (green) and $\lambda$ (red dashed) of the self-consistency equations \cite{Para1}, that show the mean-field phase transition at $T_c$. The delocalization of the $d$ electrons is described by the appearance of finite negative $\phi$ and positive $b$. The inset shows a zoom of the $T_c$ region.}
\label{Fig1}
\end{figure}

Thus, with the help of the static and uniform fields $\lambda$, $b$ and $\phi$ we have reduced the effective problem to a solvable single-particle model
\begin{eqnarray}
\mathcal{H}_{MF}= \sum_{ \bf {k}\sigma } \epsilon_{\bf {k}} c^{\dagger}_{\bf {k}, \sigma} c_{ \bf {k},\sigma} + \sum_{i \sigma} (E_0+\lambda) d^{\dagger}_{i \sigma} d_{i \sigma} -\lambda + \nonumber\\
\sum_{i}(\lambda b^2 + W \phi^2) + \sum_{i \sigma} V b d^{\dagger}_{i \sigma} c_{i, \sigma} - \sum_{i \sigma} W \phi d^{\dagger}_{i \sigma} c_{i\sigma}
\end{eqnarray}
(plus three free-fermion Hamiltonians), where we have dropped the tildes of the slave-fermion operators. The self-consistency equations for $\lambda, b$ and $\phi$ are
\begin{eqnarray}
& &V T \sum_{k, \mu, \omega_n, \sigma} G_{dc} + \lambda b=0,\\
&  &T \sum_{k, \omega_n, \sigma} G_{dc} - \phi = 0,\\
& &b^2 + T \sum_{k,  \omega_n, \sigma} G_{d} = 1,
\end{eqnarray}
where we have defined the following Green's functions:
\begin{eqnarray}
& &G_{dc}=\frac{\widetilde{V}}{(i \omega_n - \epsilon_{k})(i \omega_n - \epsilon_d)-\widetilde{V}^2},\\
& &G_{c}=\frac{i \omega_n - \epsilon_d}{(i \omega_n - \epsilon_{k})(i \omega_n - \epsilon_d)-\widetilde{V}^2},\\
& &G_{d}=\frac{i \omega_n - \epsilon_{k}}{(i \omega_n - \epsilon_{k})(i \omega_n - \epsilon_d)-\widetilde{V}^2},
\end{eqnarray}
with $\epsilon_d=E_0 + \lambda$ and $\widetilde{V} = V b - W \phi$.
We solve these equations to obtain the behavior of the mean fields with temperature, and show the result on Fig. \ref{Fig1}. As can be seen, there is a phase transition at temperature $T_c$. Above it we have $b=0$, $\phi=0$ and $\lambda=1$ -- the itinerant and the localized states are decoupled. Below $T_c$ $b$ and  $\phi$ simultaneously develop non-zero expectation value, and $\lambda$ becomes bigger than $1$; this describes the delocalization of the $d_{xy}$ states through coupling with the $d_{xz}/d_{yz}$ bands, and the opening of a hybridization gap $\widetilde{V}$ in the electronic spectrum (also, note that $\phi$ is negative -- it increases the gap, and even dominates it at low temperatures). This phase transition, however, is an artefact of the mean field treatment --  finite $b$ violates the local gauge symmetry of the model. Formally, infrared fluctuations destroy the order and restore the symmetry \cite{Read, Auerbach, Millis}. Nevertheless, as in the case of heavy-fermions we can take the mean field transition as an indication of a crossover in the real system.  

{\it Breaking of the $C_4$ symmetry.} To see how the symmetry between $\phi_3$ and $\phi_4$ appears, and how it can be broken we consider a Ginzburg-Landau (GL) theory for the fields $b$ and $\phi_{\mu}$ 
(again assuming $\phi_1=\phi_2=0$ for simplicity). 
For the symmetric case of identical circular electron bands the GL free energy takes the form:
\begin{eqnarray}
\mathcal{F}_{0}=\alpha_b b^2 + \beta_b b^4  +
 (\alpha_{\phi}|{\bf \phi}|^2 + \beta_{\phi} |{\bf \phi}|^4)  \nonumber \\ + \sum_{\mu} \gamma_{1} b|\phi_{\mu}| \cos{\varphi_{\mu}},
\end{eqnarray}
where $\varphi_{\mu}$ is the phase of $\phi_{\mu}$. The coefficients in this (bosonic) theory can be obtained by integrating out the fermions, and depend implicitly on $\lambda$.

Now it is clear why we had to take the the symmetric combination of $\phi_3$ and $\phi_4$ -- $b$ acts as an external field in the $(1, 1)$ direction of the $(\phi_3, \phi_4)$ space. Thus the $C_4$ symmetry is preserved both above and below the transition. The linear coupling also guarantees that $b$ and $\phi$ become non-zero simultaneously, and justifies our choice of mean-field decoupling of the $W$ term.

In realistic models the symmetries of the band structure are significantly reduced. First, the hole pockets have different $k_F$ (thus $\phi_1$ and $\phi_2$, if present, are not degenerate). Second, the electron pockets become ellipses (but are still identical after translation and $\pi/2$ rotation). In terms of orbital content the $c_1$ and $c_2$ states are still equal mixtures of $d_{xz}$ and $d_{yz}$ (averaged over their respective Fermi surfaces). In contrast, the $c_3$ and $c_4$ are dominated by either one or the other orbital. Importantly, this means that any $\phi$ order parameter that is not $3 \leftrightarrow 4$ symmetric will simultaneously break the $C_4$ symmetry of the lattice, and induce orbital order.

\begin{figure}[h]
\begin{center}$
\begin{array}{cc}
\includegraphics[width=0.23\textwidth]{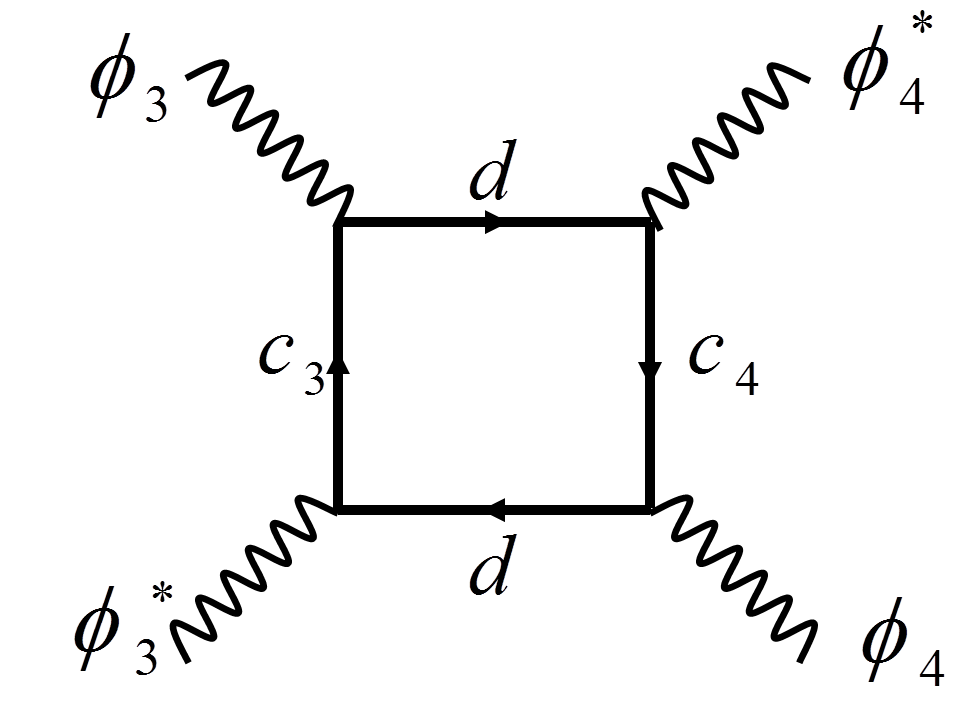}
\includegraphics[width=0.23\textwidth]{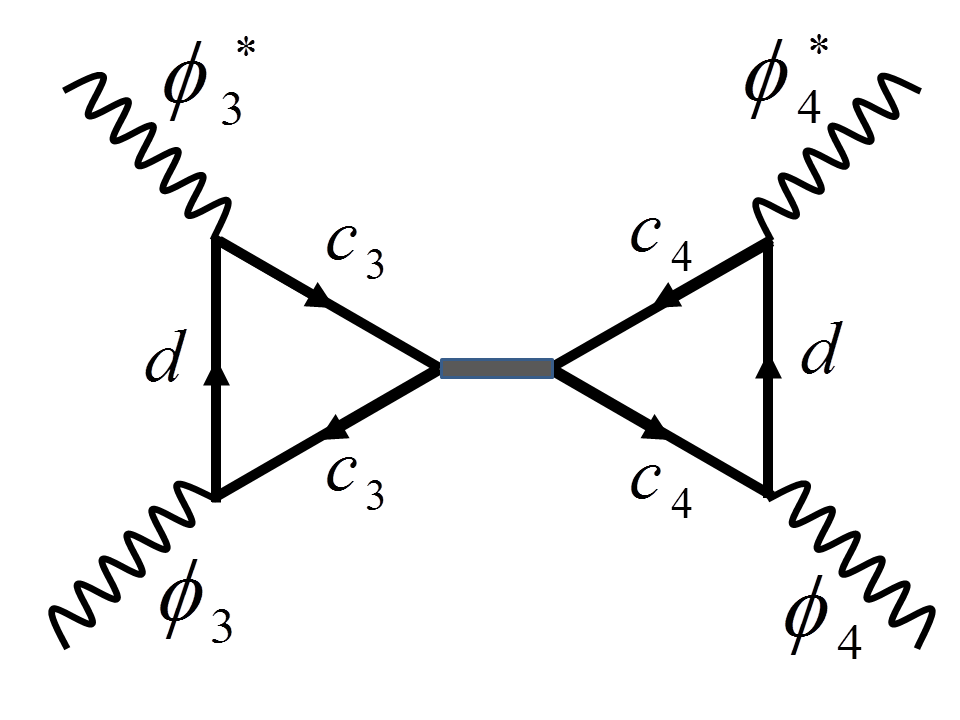}
\end{array}$
\end{center}
\caption{On the left is the Feynman diagram that contributes to $\gamma_2$ because of the ellipticity of the $c_3$ and $c_4$ bands. On the right is diagram of a process generated by the interband interactions.}
\label{Fig2}
\end{figure}

It is easy to see that the ellipticity of the bands leads to additional quartic term in the effective action:
\begin{eqnarray}
\mathcal{F}'=  \sum_{\mu, \nu} \gamma_2|\phi_\nu|^2|\phi_\mu|^2,
\end{eqnarray}
with $\mu \neq  \nu$, through the standard four-fermion diagram on the left of Fig. \ref{Fig2}. The value and the sign of the coefficient depend on the ellipticity, but it can be shown by direct calculation that for parabolic bands with small ellipticity $\gamma_2$ is positive. The physical origin of this correction is the particle-hole asymmetry.

There is yet another source of quartic terms -- the renormalized interactions in the $d_{xz}/d_{yz}$ subspace hidden in $\mathcal{H}_{res}$. 
In particular, $c^{\dagger}_{\mu} c_{\mu} c^{\dagger}_{\nu} c_{\nu}$ term generate processes like the one shown on the right of Fig. \ref{Fig2}. They also contribute to $\gamma_2$, and again it can be easily shown that their contribution is positive. Note that these are interband interactions terms, which prefer the delocalized $d$ electrons to mix with $c_3$ \emph{or} $c_4$; the (presumably larger) intraband terms participate in setting the overall scale the order parameter, but do not contribute to $\gamma_2$.

\begin{figure}[h]
\begin{center}$
\begin{array}{cc}
\includegraphics[width=0.45\textwidth]{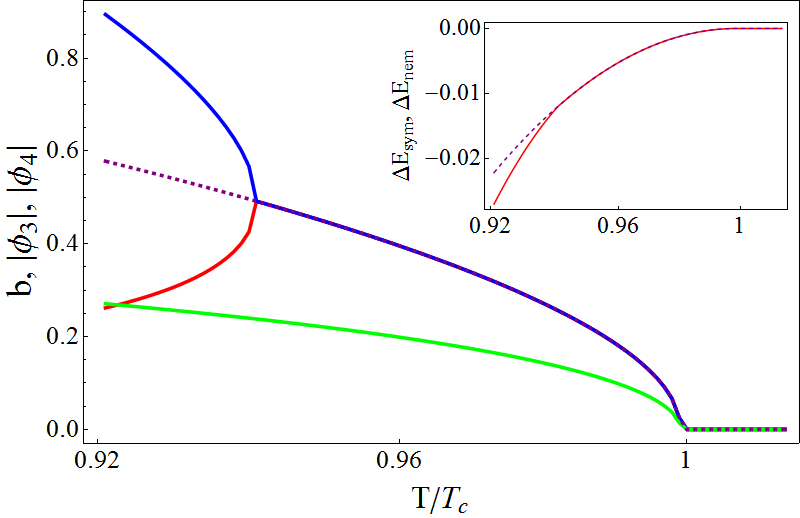}
\end{array}$
\end{center}
\caption{Plot of $b$ (green), $\phi_3$ (red) and $\phi_4$ (blue) that minimize the GL functional \cite{Para2}. Below $T\approx 0.94 T_c$ the effect of positive $\gamma_2$ overcomes the linear coupling with $b$ and additional solution appears, with $\phi_3\neq \phi_4$. In the inset we show that the free energy of this solution (red) is below the free energy of the symmetric solution (purple). The new order parameter breaks the $C_4$ symmetry.}
\label{Fig3}
\end{figure}

 If $\gamma_2$ is positive it penalizes the coexistence of $\phi_3$ and $\phi_4$, and favors an order parameter with only one non-zero component. The effect of this term, however, can overcome the linear coupling of $\phi_3$ and $\phi_4$ to $b$ only at a finite temperature below $T_c$; this is demonstrated on Fig. \ref{Fig3}, in which we show solutions of the GL equations below $T_c$.  We see that at temperatures sufficiently below $T_c$ an asymmetric combination of $\phi_3$ and $\phi_4$ becomes a minimum of the free energy. This state breaks the $C_4$ symmetry of the underlying lattice and reduces it to $C_2$. 

Thus in our model the nematic state emerges from a two-stage process -- at the first stage the $d_{xy}$ electrons, localized at high temperatures, start to mix and form coherent bands with the itinerant electrons. The quasiparticles that emerge at this stage are strongly renormalized by local interactions like Hubbard's $U$ and Hund's $J_H$. Note the parallels not only to the PAM, but also to the Zaanen-Sawatzky-Allen work \cite{Zaanen}, in which $d$-orbital electrons delocalize through coupling with valence bands.  At a lower temperature, possibly far below the mean-field transition, the order parameter breaks the lattice rotation symmetry. This is the only true phase transition of the model, and it is driven by band-structure details and the interband interactions of the itinerant electrons. Those are relatively weak effects, compared to the local physics dominating at higher temperatures, and they can play a nontrivial role only because the first transition (actually crossover) has significantly reduced the effect of the strong local interactions on the low-energy physics.

In this scenario the nematic state is not induced by proximity to the antiferromagnetic state, but is an independent instability. The presence of nematicity, however, can, by itself, enhance the magnetic instability \cite{WCL2}. Also, note the similarities of our model with the physics of some nickel-based compounds \cite{Ni1, Ni2}. 

{\it Conclusion.} In this paper we introduced an effective model for the normal state of iron-based superconductors. It has both itinerant and localized degrees of freedom -- the former originate from the $d_{xz}/d_{yz}$, and the latter from $d_{xy}$ iron orbitals. We studied this model on a mean field level and showed that at low temperatures the $d_{xy}$ states can effectively delocalize and condense together with the itinerant states in an excitonic order parameter.  Because of the multiband character of the itinerant Fermi surface the coupling between the localized and itinerant electrons can naturally lead to another phase transition, and a nematic excitonic state. We propose this mechanism as an explanation of the tendency towards nematicity observed in several iron-based compounds.

{\it Acknowledgements}. We gratefully acknowledge insightful discussions with Rebecca Flint, Jasper van Wezel and Zlatko Te\v sanovi{\' c}. This work was supported by U.S. DOE, Office of Basic Energy Sciences, under contract no. DE-AC02-06CH11357, and the Center for Emergent Superconductivity, a DOE Energy Frontier Research Center, under contract no. DE-AC0298CH1088.

 \bibliographystyle{apsrev}

\begin{thebibliography}{10}
\bibitem{Para1} For the plot we have used $E_0=-1$, $W=1$, $V=1$, energy cut-off $\Lambda=100$ and density of states $N=1$ as parameter values.
\bibitem{Para2} For the plot we have used $\alpha_b=(T-T_c)$, $\alpha_{\phi}=(T-T_c + 0.3)$, $\beta_b=0.1$, $\beta_{\phi}=0.1$, $\gamma_1=0.2$, $\gamma_2=0.2$.
\end{thebibliography}

\end{document}